\documentclass[a4paper,twoside,12pt]{article}
\input{obsjkt.sty}

\newcommand{\Msun}{\ensuremath{~{\rm M}_\odot}}                   
\newcommand{\Rsun}{\ensuremath{~{\rm R}_\odot}}                   
\newcommand{\Teff}{\ensuremath{T_{\rm eff}}}                      
\newcommand{\logg}{\ensuremath{\log g}}                           
\newcommand{\degr}{\ensuremath{^\circ}}                           
\newcommand{\mc}[1]{\multicolumn{2}{c}{#1}}
\newcommand{\etal}{\textit{et al.}}                               
\newcommand{\corot}{\textit{CoRoT}}
\newcommand{\kepler}{\textit{Kepler}}
\newcommand{\tess}{\textit{TESS}}

\usepackage{rotating}

\begin{document} 

\OBSheader{Rediscussion of eclipsing binaries: V498 Cygni}{J.\ Southworth}{2021 Dec}

\OBStitle{Rediscussion of eclipsing binaries. Paper VIII. \\ The doubly-eclipsing quadruple star system V498 Cygni}

\OBSauth{John Southworth}

\OBSinstone{Astrophysics Group, Keele University, Staffordshire, ST5 5BG, UK}


\OBSabstract{V498~Cyg is an early-B-type binary known to show eclipses on a period of 3.48~d, and two sets of spectral lines. We present the discovery of a second set of eclipses, on a 1.44-d period, in the light curve of this object from the \textit{Transiting Exoplanet Survey Satellite} (\tess). We develop a model of the light curve to simultaneously fit the properties of both eclipsing binaries and apply this to the \tess\ observations. We are able to fit the light curve of the fainter system well, but the light curve fit for the brighter system is unable to reproduce either its asymmetric primary eclipse or its changing light curve shape. The available eclipse timing measurements are extremely scattered so we determine orbital ephemerides based only on the \tess\ data. We infer the physical properties of all four stars, estimating the masses of the components of the brighter binary to be 10\Msun\ and 11\Msun, and of the fainter binary to be 6.5\Msun\ and 3.5\Msun. The properties of the system may be reliably determined in future by obtaining radial velocity measurements of the component stars.}


\section*{Introduction}

Detached eclipsing binary stars (dEBs) are a vital source of directly-measured properties of normal stars against which our understanding of stellar physics can be examined \cite{Andersen++90apj,Torres++10aarv,Me20obs}. High-mass stars are known to be found predominantly in systems with a high incidence of binarity or higher-order multiplicity \cite{DucheneKraus13araa,Sana+14apjs}. The binary fraction, plus the distributions of mass ratio and orbital eccentricity, is useful in understanding the formation processes of single and binary stars \cite{Bate09mn,Bate19mn,Moe++19apj,JustesenAlbrecht21apj}.

As a small fraction of stellar systems are quadruple or of higher multiplicity, it is possible to observe two dEBs in one system. Although a few have been discovered from ground-based observations (V994~Her \cite{Lee+08mn}, CZeV343 \cite{CagasPejcha12aa}, V482~Per \cite{Torres+17apj}), the great majority have only recently been identified using data from the current generation of space-based photometric surveys such as \kepler, \corot\ and \tess, including  KIC 4247791 \cite{Lehmann+12aa}, KIC 4150611 \cite{Helminiak+17aa}, EPIC 219217635 \cite{Borkovits+18mn}, TIC 278956474 \cite{Rowden+20aj}, BG~Ind \cite{Borkovits+21mn}, TIC 454140642 \cite{Kostov+21apj} and BU~CMi \cite{Volkov++21}. The system TIC 168789840 has been found to contain \emph{three} dEBs with short orbital periods (1.57, 8.21 and 1.31 d) and strikingly similar primary (each 1.2--1.3\Msun) and secondary (each approximately 0.6\Msun) components. A detailed review of the impact of space-based photometry on binary star science can be found in Southworth \cite{Me21univ}.

As the multiplicity of massive stars is higher, it is to be expected that some such systems might host two or more eclipsing binaries. However, to our knowledge, the earliest spectral type found in the systems mentioned above is B8\,V. In this work we present the discovery that V498~Cyg shows two sets of eclipses, on periods of 3.48 and 1.44 d. The two stars in the brighter dEB are of an early-B spectral type, making this system possibly the most massive doubly-eclipsing binary known.


\begin{table}[t]
\caption{\em Basic information on V498~Cyg \label{tab:info}}
\centering
\begin{tabular}{lll}
{\em Property}                   & {\em Value}            & {\em Reference}                \\[3pt]
Henry Draper designation         & HD 229179              & \cite{Cannon36anhar}           \\
\textit{Tycho} designation       & TYC 3152-577-1         & \cite{Hog+00aa}                \\
\textit{Gaia} EDR3 designation   & 2061233871414590208    & \cite{Gaia21aa}                \\
\textit{Gaia} EDR3 parallax      & $0.591 \pm 0.016$ mas  & \cite{Gaia21aa}                \\
\tess\ designation               & TIC 13968858           & \cite{Stassun+19aj}            \\
$B_T$ magnitude                  & $10.934 \pm 0.057$     & \cite{Hog+00aa}                \\
$V_T$ magnitude                  & $9.977 \pm 0.029$      & \cite{Hog+00aa}                \\
$J$ magnitude                    & $7.657 \pm 0.019$      & \cite{Cutri+03book}            \\
$H$ magnitude                    & $7.447 \pm 0.017$      & \cite{Cutri+03book}            \\
$K_s$ magnitude                  & $7.318 \pm 0.020$      & \cite{Cutri+03book}            \\
Spectral type                    & B1.5 V                 & \cite{Lacy90ibvs}              \\[10pt]
\end{tabular}
\end{table}

\section*{V498 Cygni}

Some basic information on V498~Cyg is given in Table~\ref{tab:info}. The $B_T$ and $V_T$ magnitudes are from the Tycho-2 catalogue \cite{Hog+00aa} and are the averages of thousands of measurements covering all orbital phases. The $JHK_s$ magnitudes are single-epoch and their orbital phases cannot be established precisely (see below) so should be read as indicative values only.

The literature on V498~Cyg is sparse. Hiltner \cite{Hiltner56apjs} assigned a spectral classification of B1:III:, inconveniently given with designation BD +38\degr4054 and its HD number missing from the relevant column. The variability of the system was discovered by Hoffmeister \cite{Hoffmeister36an} and described as short-period. Quite a few studies have presented times of minimum light or light curves of often limited quality \cite{Sandig48an,Einasto51pz,Romano69mmsai,MagalashviliKumsishvili78abaob,Mayer+91baicz,SmithCaton07ibvs,HubscherLehmann15ibvs}.

The only substantive analysis of the system is by Zakirov \& Eshankulova \cite{ZakirovEshankulova08kpcb}, who presented extensive photometry in the Johnson $UBVR$ system. They found clear evidence for orbital eccentricity: the secondary eclipse occurred at phase 0.526 and its duration differed from that of the primary (0.18 versus 0.23 in phase units). They also found that ``the light curves have some irregularities and the shape of the curves is unstable in all spectral bands'' and tentatively attributed this to the presence of a third body.

V498~Cyg originally came to the author's attention due to a paper by Lacy \cite{Lacy90ibvs} which reported the discovery of two sets of spectral lines. Lacy assigned a spectral type of B1.5\,V to the system and stated that the depths and widths of the lines from the two stars were similar. The dwarf spectral classification is more consistent with the expected sizes of the stars (the system is detached and has a 3.48~d period) than the previous classification as a giant, so should be preferred. V498~Cyg was included in a systematic trawl through a list of interesting eclipsing binaries with light curves in the TESS database, and it was immediately noticed that two sets of eclipses were present. The analysis of these data is reported below. Whilst this work was being performed, Eisner \etal\ \cite{Eisner+21mn} independently discovered the multi-eclipsing nature of the system but did not present any analysis.


\section*{Observational material}

\begin{figure}[t] \centering \includegraphics[width=\textwidth]{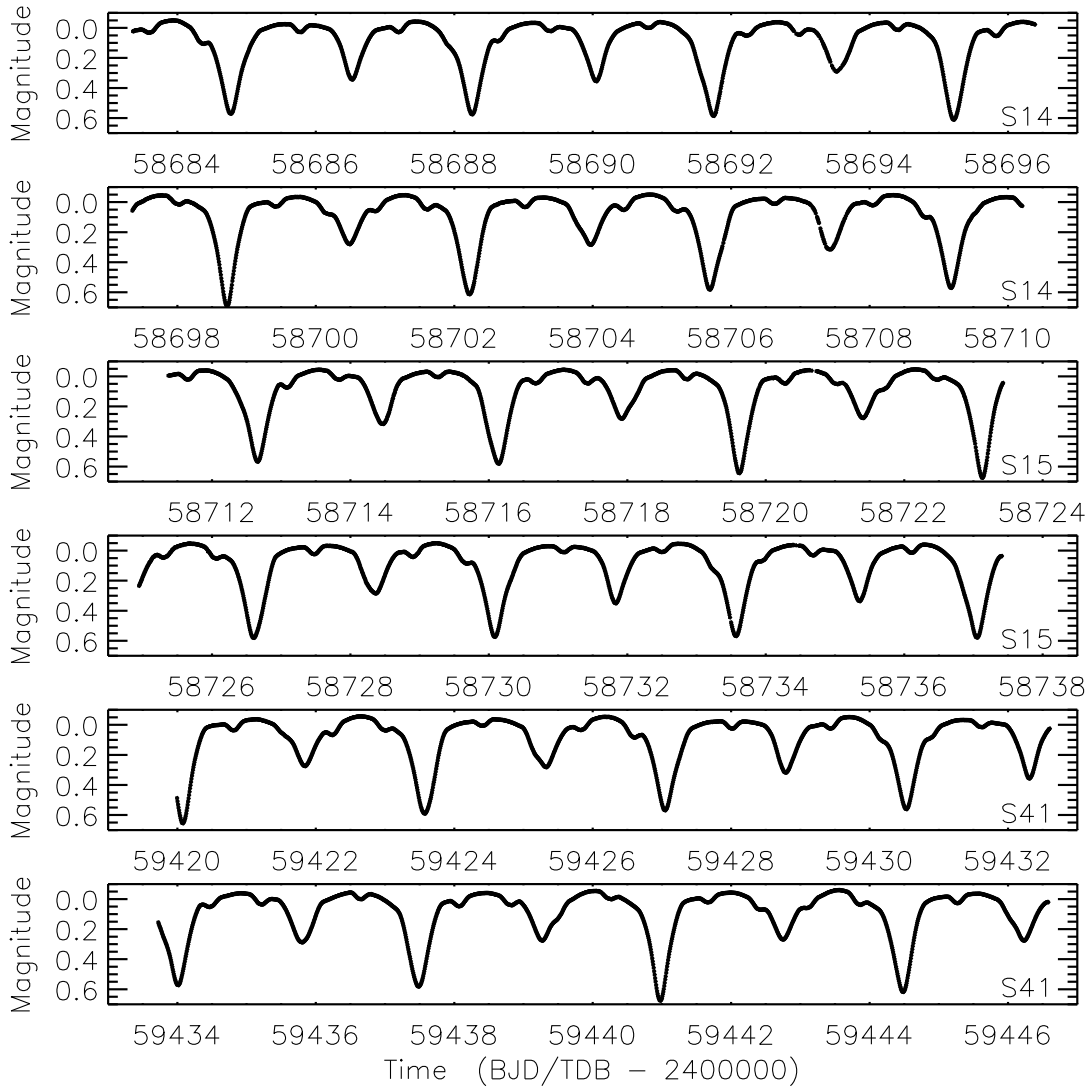} \\
\caption{\label{fig:time} \tess\ short-cadence SAP photometry of V498~Cyg from sectors 14, 15 and 41 (labelled).
Each panel contains half a sector of data, from either before or after the mid-sector gap in observations. The
flux measurements have been converted to relative magnitude and rectified to zero magnitude by subtraction of
low-order polynomials. The observations have also been binned by a factor of five before plotting to decrease
the size of the image file; this has a negligible effect on the appearance of the figure.} \end{figure}


V498~Cyg has been observed on three occasions by the NASA \tess\ satellite \cite{Ricker+15jatis}: in sectors 14 (2019/07/18 to 2019/08/15), 15 (2019/08/15 to 2019/09/11) and 41 (202107/23 to 2021/08/20); and a fourth visit is planned (sector 55 in 2022 August). In each case the observations lasted for approximately 27~d and were continuously sampled at a cadence of 120~s except for mid-sector pauses for the data to be downlinked to Earth.

The data were downloaded from the MAST archive\footnote{Mikulski Archive for Space Telescopes, \\ \texttt{https://mast.stsci.edu/portal/Mashup/Clients/Mast/Portal.html}} and converted to relative magnitude. Observations with a QUALITY flag of less than 5000 were retained as they appeared to be reliable; restricting to a QUALITY of zero would cause the loss of several thousand datapoints. We used the simple aperture photometry (SAP) version of the \tess\ data \cite{Jenkins+16spie} as it is the more suitable for stars with strong intrinsic brightness variations. The light curves are shown in Fig.~\ref{fig:time}.


\section*{A model for the \tess\ light curve}

\begin{figure}[t] \centering \includegraphics[width=\textwidth]{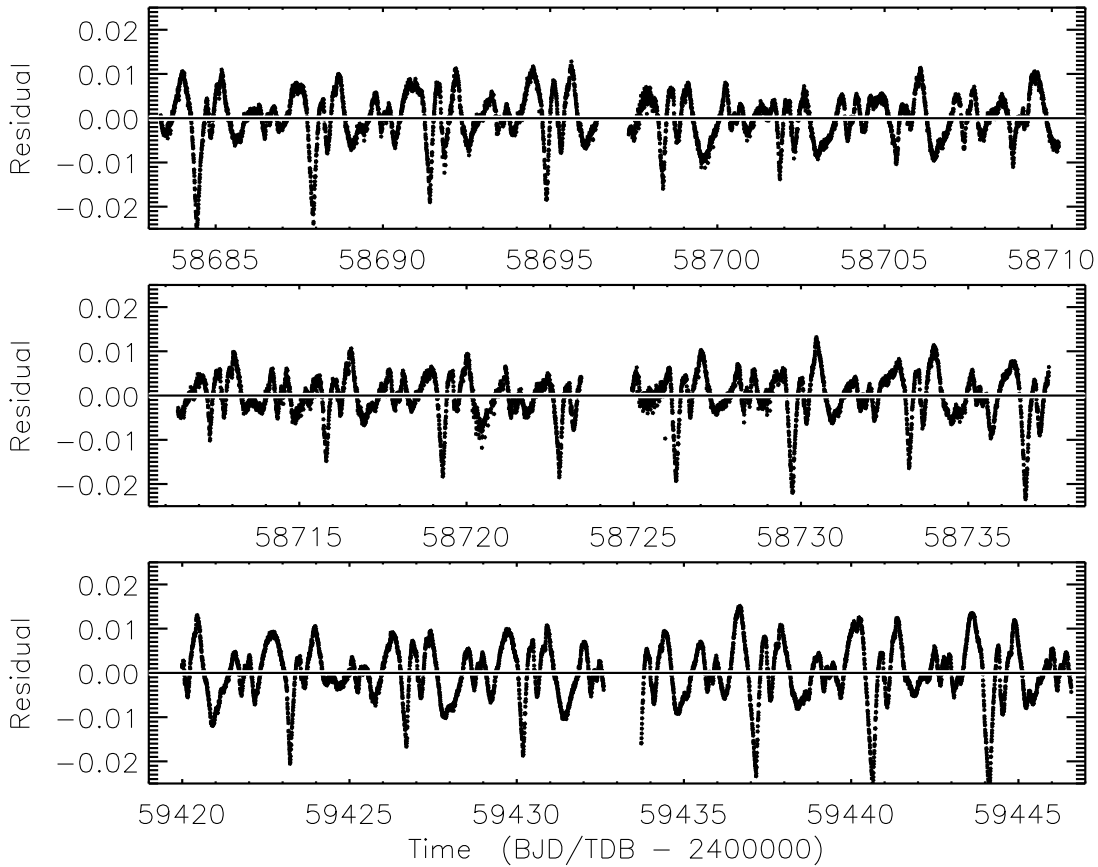} \\
\caption{\label{fig:resid} Residuals of the best fits to the \tess\ SAP data in
sectors 14 (top panel), 15 (middle panel) and 41 (bottom panel).} \end{figure}

The two sets of eclipses shown by V498~Cyg dictated the development of a bespoke model for the light curve of the system. From this point onwards we refer to the brighter of the dEBs as binary~A, which contains individual stars Aa and Ab where Aa is the primary star and has a higher surface brightness than the secondary star Ab. Similarly, the fainter dEB is referred to as binary~B and the individual stars as Ba and Bb. The model of the system comprises the light contributions from the two dEBs, a third light and a quadratic function to scale the overall light curve to the observed magnitudes. This is expressed  according to the equation
\begin{equation}
\ell_{\rm total}(t) = \frac{\ell_{\rm A}(t) + \alpha\ell_{\rm B}(t) + \ell_3}{1+\alpha+\ell_3} \Big[p_1+p_2(t-t_{\rm piv})+p_3(t-t_{\rm piv})^2\Big]
\end{equation}
where $\ell_{\rm total}(t)$ is the total light from the system. $\ell_{\rm A}(t)$ and $\ell_{\rm B}(t)$ are the light curves of the two dEBs, each normalised to unit flux at quadrature. $\ell_3$ is the time-independent contaminating `third' light from any other bodies that may contribute to the light curve, expressed as a fraction of the total light from the system. The light ratio between the two dEBs is given by $\alpha$ and is the ratio of the brightness of binary~B to that of binary~A, each taken at quadrature. Finally, the quadratic polynomial has coefficients $p_1$, $p_2$ and $p_3$ and is pivoted around time $t_{\rm piv}$ for numerical stability, where $t_{\rm piv}$ is a point near the centre of the data under consideration.

\begin{figure}[t] \centering \includegraphics[width=\textwidth]{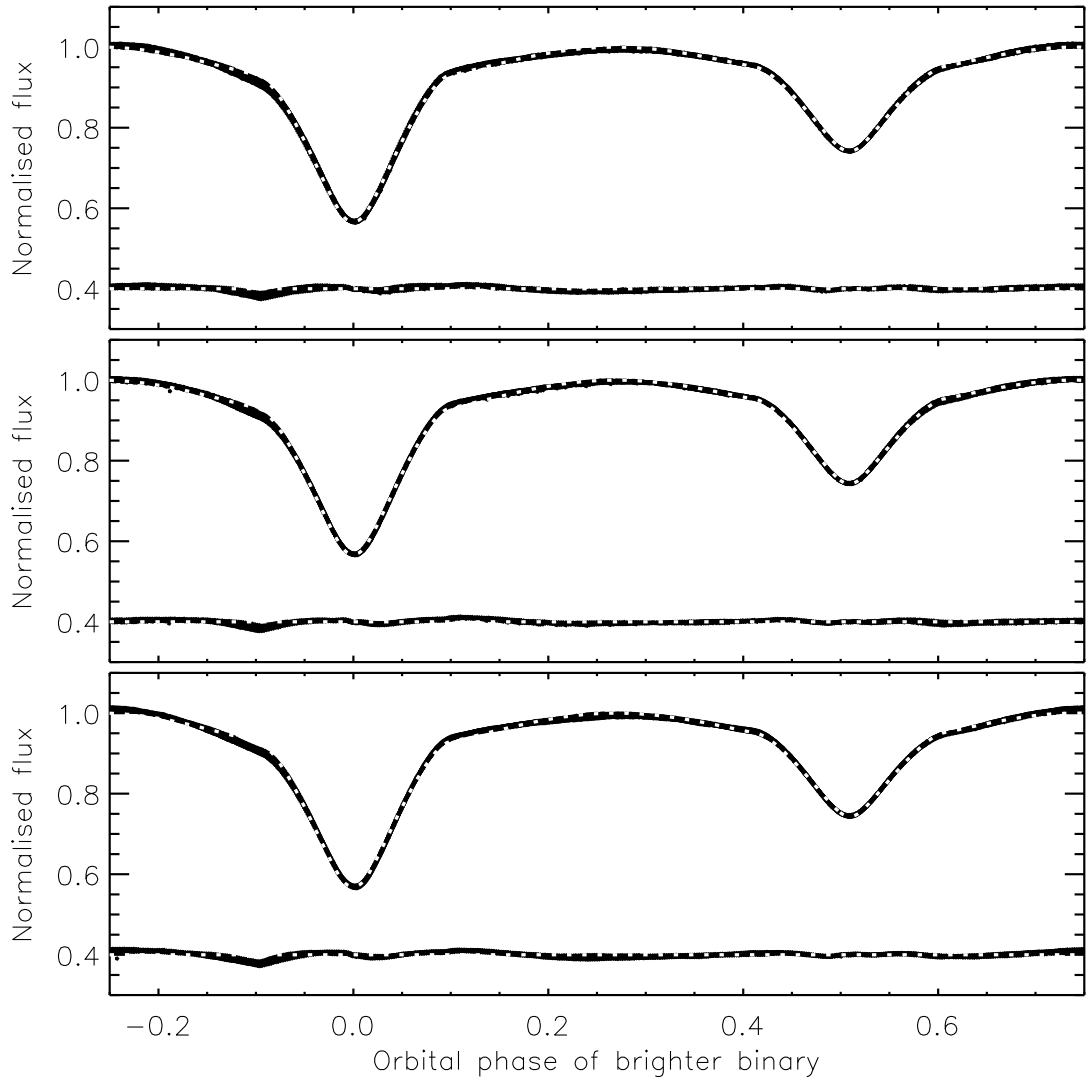} \\
\caption{\label{fig:phase1} Best fit to the \tess\ SAP data in sectors 14 (top panel),
15 (middle) and 41 (bottom) for binary~A. In each panel the \tess\ data are shown after
subtraction of the model light curve of binary~B, third light and the quadratic function.
The best fits are shown with white dotted lines superimposed on the data. The residuals
are shown at the base of each panel, offset from zero.} \end{figure}

To calculate $\ell_{\rm A}(t)$ and $\ell_{\rm B}(t)$ we used the {\sc jktebop}\footnote{\texttt{http://www.astro.keele.ac.uk/jkt/codes/jktebop.html}} code \cite{Me++04mn2,Me13aa}. The fitted parameters for each dEB included the sum and ratio (primary divided by secondary) of their fractional radii, the orbital inclination and period, a reference time of midpoint of the primary eclipse, and the central surface brightness ratio of the two stars (primary divided by secondary). We adoped the quadratic limb darkening law but fixed the coefficients to reasonable values from theoretical predictions \cite{Claret18aa}. A circular orbit was adequate for binary~B but an orbital eccentricity was required for binary~A. We therefore fitted for the parameters $e\cos\omega$ and $e\sin\omega$, where $e$ is the orbital eccentricity and $\omega$ is the argument of periastron, for binary~A. We also fitted for $\alpha$, $\ell_3$, and the polynomial coefficients.

\begin{figure}[t] \centering \includegraphics[width=\textwidth]{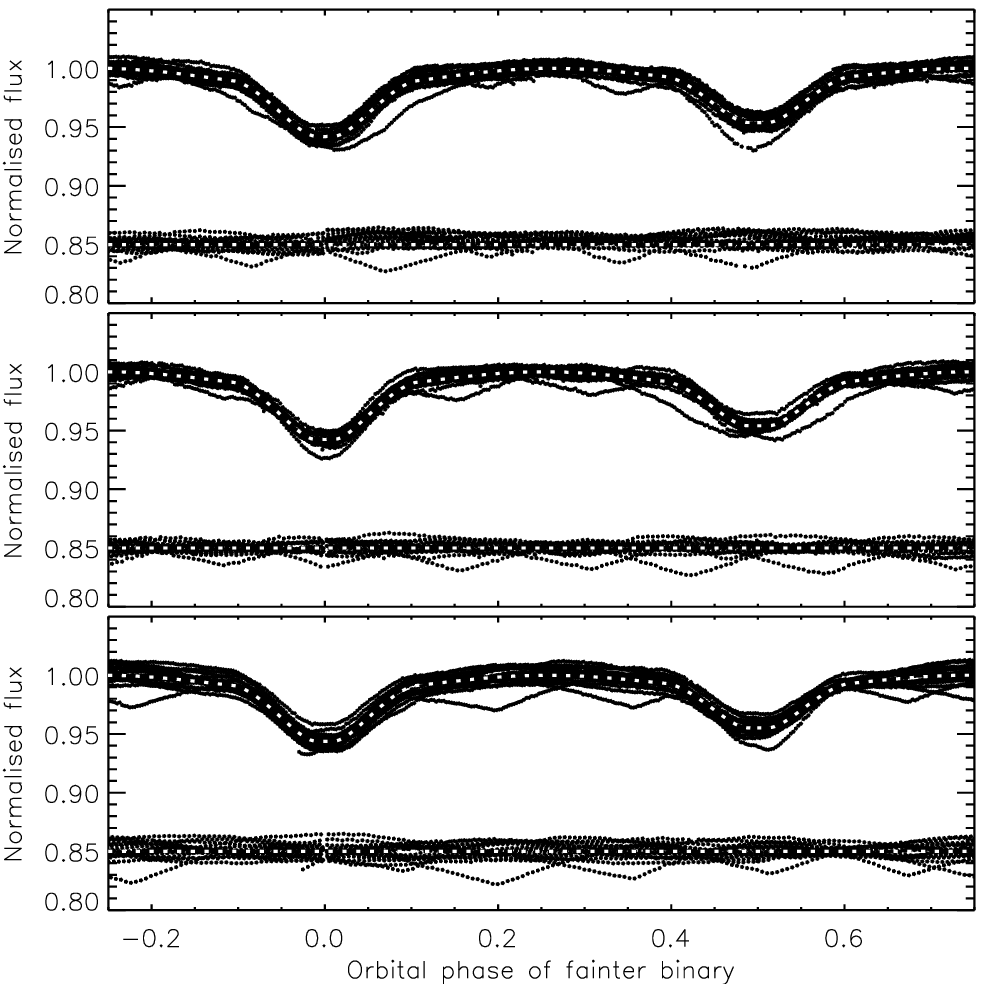} \\
\caption{\label{fig:phase2} Best fit to the \tess\ SAP data in sectors 14 (top panel),
15 (middle) and 41 (bottom) for binary~B. In each panel the \tess\ data are shown after
subtraction of the model light curve of binary~A, third light and the quadratic function.
The best fits are shown with white dotted lines superimposed on the data. The residuals
are shown at the base of each panel, offset from zero.} \end{figure}

This fitting code was named {\sc multebop} and implemented in an IDL\footnote{Interactive Data Language, \\ \texttt{https://www.l3harrisgeospatial.com/Software-Technology/IDL}} script that called {\sc jktebop} to generate model light curves and optimised the parameters of the model using the Levenberg-Marquardt method \cite{Marquardt63} as implemented in the {\sc mpfit} package \cite{Markwardt07aspc}. Exploratory solutions were performed with the data from a single sector, binned by a factor of three, for speed. The errorbars of individual datasets were scaled to yield a reduced $\chi^2$ of $\chi^2_\nu = 1$.

Attempts to fit for the limb darkening coefficients typically returned very similar parameters but fitted coefficients very different from the expected values and sometimes unphysical (i.e.\ not within the interval [0,1]). The third light, however, is reliably detected. It could be caused by (yet) another stellar component in the system, imperfections in the {\sc multebop} model, and/or inaccurate background subtraction in the reduction of the \tess\ data. The presence of a fifth star is the most plausible answer as the third light is approximately 10\% of the total light of the system so is larger than expected for the other two explanations.

The residuals of the best fits are shown in Fig.~\ref{fig:resid}, and the best fits to the individual sectors are shown in Fig.~\ref{fig:phase1} for binary~A and Fig.~\ref{fig:phase2} for binary~B. The orbital ephemerides are not reported as they will be discussed in a following section. It is clear that we are unable to obtain a good fit to these data, and this problem is most apparent in our inability to properly fit the data around the start and end of primary minimum in binary~A (Fig.~\ref{fig:phase1}). This poor fit is the primary driver for systematic variations between orbital cycles visible for binary~B (Fig.~\ref{fig:phase2}). There are two problems here.

First, the light variation of binary~A changes on the timescale of several orbital cycles: this can be seen in the varying structure of the residuals in Fig.~\ref{fig:resid}, especially during sectors 14 and 41. This effect is conceivably due to pulsations in one or both components, as they are in the region of the HR diagram where $\beta$~Cephei pulsations are frequently found \cite{Walczak+15aa,Me+20mn,Me++21mn}, or to hot spots or gas streams caused by a low level of mass transfer or stellar winds. The light curve model is not able to produce a significant asymmetry in primary eclipse without extreme assumptions on orbital eccentricity or the reflection effect, so cannot adequately trace the true light variation in the data.

\begin{table} \centering
\caption{\em \label{tab:lc} Best-fitting parameters and uncertainties from the {\sc multebop} fit to the
three sectors of data. The parameter values are the means from fits to the three sectors individually. The
uncertainties are the standard deviations (not standard errors) of the fits to six half-sectors individually.}
\begin{tabular}{lr@{\,$\pm$\,}lr@{\,$\pm$\,}l}
{\em Parameter}                             & \multicolumn{2}{c}{\em Binary A} & \multicolumn{2}{c}{\em Binary B} \\[3pt]
{\it Fitted parameters:} \\
Orbital inclination (\degr)                 &      86.96      & 0.20            &      83.7       & 1.6             \\
Sum of the fractional radii                 &       0.5705    & 0.0023          &       0.6201    & 0.0023          \\
Ratio of the radii                          &       1.2145    & 0.0067          &       0.606     & 0.032           \\
Central surface brightness ratio            &       0.5303    & 0.0028          &       0.730     & 0.064           \\
Linear LD coefficient for primary star      & \multicolumn{2}{c}{0.05 (fixed)}  & \multicolumn{2}{c}{0.08 (fixed)}  \\
Quadratic LD coefficient for secondary star & \multicolumn{2}{c}{0.05 (fixed)}  & \multicolumn{2}{c}{0.08 (fixed)}  \\
Linear LD coefficient for primary star      & \multicolumn{2}{c}{0.24 (fixed)}  & \multicolumn{2}{c}{0.21 (fixed)}  \\
Quadratic LD coefficient for secondary star & \multicolumn{2}{c}{0.24 (fixed)}  & \multicolumn{2}{c}{0.21 (fixed)}  \\
$e\cos\omega$                               &       0.01439   & 0.00061         & \multicolumn{2}{c}{0.0  (fixed)}  \\
$e\sin\omega$                               &    $-$0.0086    & 0.0065          & \multicolumn{2}{c}{0.0  (fixed)}  \\
Light ratio between dEBs ($\alpha$)         &                 \multicolumn{4}{c}{$0.203 \pm 0.018$}                 \\
Third light                                 &                 \multicolumn{4}{c}{$0.096 \pm 0.016$}                 \\[5pt]
{\it Derived parameters:} \\
Fractional radius of primary star           &       0.2576    & 0.0013          &       0.3861    & 0.0078          \\
Fractional radius of secondary star         &       0.3129    & 0.0015          &       0.2340    & 0.0077          \\
Orbital eccentricity                        &       0.0167    & 0.0032          &       \multicolumn{2}{c}{0.0}     \\
Argument of periastron (\degr)              &      31         & 15              &     \multicolumn{2}{c}{ }         \\
Light ratio                                 &       0.782     & 0.010           &       0.268     & 0.037           \\
\end{tabular}
\end{table}

Second, the stars in \emph{both} dEBs are significantly tidally distorted and therefore are beyond the limits of applicability of the spherical approximation implemented in {\sc jktebop}. The results in Table~\ref{tab:lc} should therefore be interpreted as indicative of the properties of the system, and not as reliable measurements of its physical properties.

Although we did not get a good fit to the \tess\ data, our individual fits of the three sectors returned parameter values that are highly consistent. For each parameter we took its final value to be a straight mean of the three values from the three sectors rather than fitting them all together. This is to guard against the possibility of orbital evolution and thus changing orbital periods and phases between sectors. To determine the uncertainties of the parameters we ran six separate fits, each corresponding to half of a \tess\ sector (split at the mid-sector gap for data download to Earth) and determined the standard deviation of the six values for each parameter. This process was intended to provide reliable errorbars whilst avoiding the computational expense of Monte Carlo methods\footnote{A single fit to one sector of \tess\ data after binning by a factor of 3 typically took approximately 6--7 minutes on the author's work laptop (a Lenovo Thinkpad L14 with Intel i5 CPU, 32GB RAM and 2TB SSD, with Kubuntu 20.04 LTS as the operating system).}. The resulting parameters and uncertainties are given in Table~\ref{tab:lc}. We checked for convergence problems by running multiple solutions from a range of starting parameter values, finding that all of them converged on the same $\chi^2$ minimum as the one corresponding to the results in Table~\ref{tab:lc}. We did not convert the standard deviations to standard errors as our imperfect fit to the data mean the parameter values are not reliable.


\section*{Separate fits to the light curves of the two dEBs}

We attempted to obtain a better fit to the light curve of binary~A using a code incorporating Roche geometry. We took the \tess\ light curve from sector 14, subtracted the fitted polynomials and model for binary~B, converted to orbital phase, and binned into 250 points equally spaced in phase. A fit was performed to this binned light curve using the Wilson-Devinney (WD) code \cite{WilsonDevinney71apj,Wilson79apj} driven by the {\sc jktwd} wrapper \cite{Me+11mn} and using the approach described in Paper~1 of this series \cite{Me20obs}. The same problem as before was found: the light curve is asymmetric around the primary eclipse (most easily seen in the differing light level immediately before and after the eclipse) and this could not be matched by us using the WD code. The solution is very sensitive to the mass ratio specified, and all attempts to fit for this parameter were terminated by a failure of the least-squares minimisation algorithm. As an improved fit to the light curve was not found during this process, we do not present any fitted parameters or a plot of the best fit.

\begin{figure}[t] \centering \includegraphics[width=\textwidth]{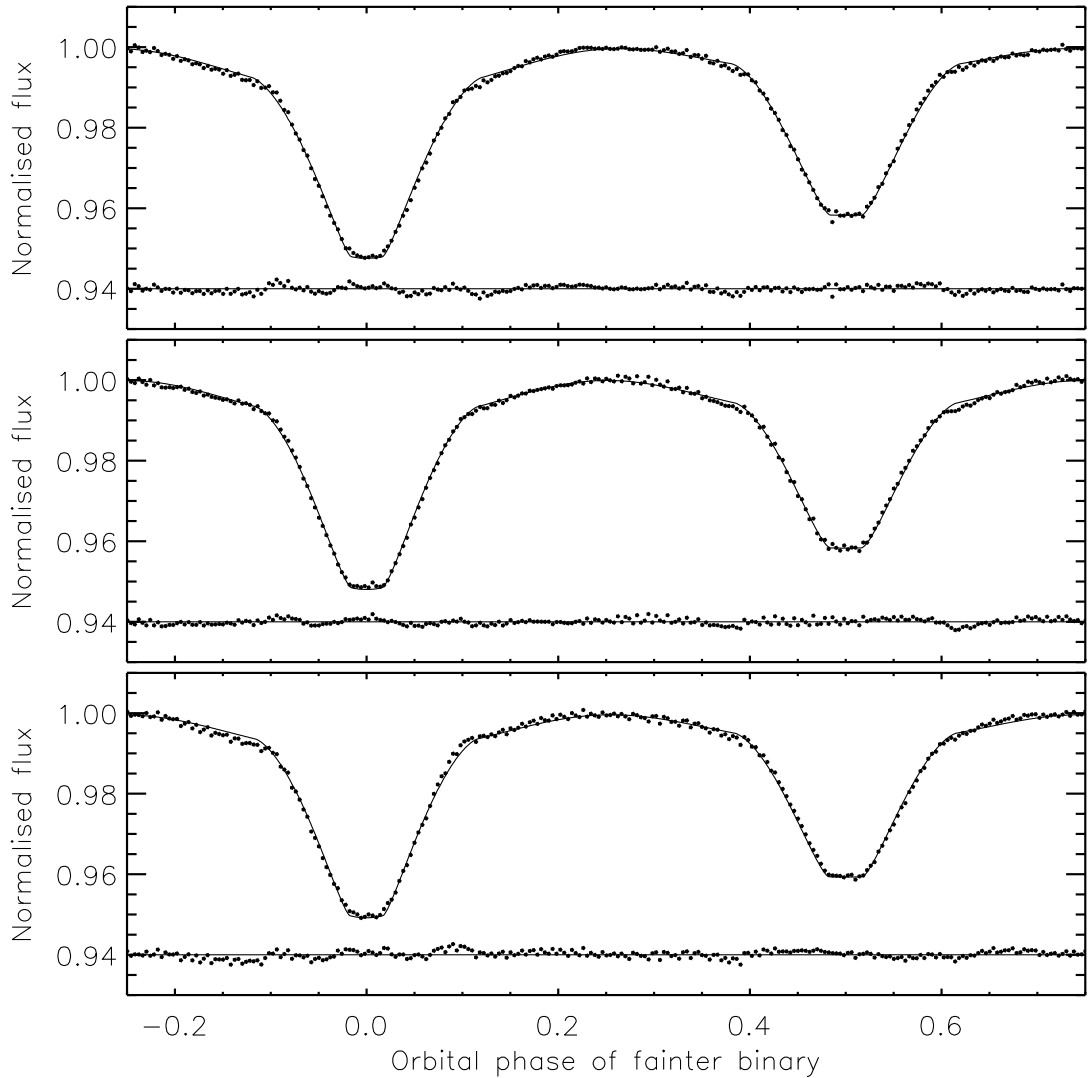} \\
\caption{\label{fig:wd2} Best fit to the \tess\ SAP data in sectors 14 (top panel), 15 (middle)
and 41 (bottom) for binary~B. The model light curve of binary~A, third light and the quadratic
function have been subtracted from the data, which has then been phase-binned using the orbital
ephemeris of binary~B. The best-fitting models from the WD code are shown with solid lines.
The residuals are shown at the base of each panel, offset from zero.} \end{figure}

A similar approach to the light curve of binary~B conversely yielded a good fit to the data. This light curve is ``better behaved'' in the sense that it more closely conforms to the expected shape for a binary system lacking mass transfer or asymmetries in the surface brightnesses of the component stars. The parameters of the fit are more reliable than those obtained using {\sc jktebop}, due to the use of Roche geometry rather than approximating the stars as spheres. Due to this relative success the same treatment was also applied to the light curves from sectors 15 and 41. The best fits are shown in Fig.~\ref{fig:wd2}, where the shapes of the total eclipses are much easier to discern than in Fig.~\ref{fig:phase2}. The fitted parameters are given in Table~\ref{tab:wd2}, where the parameter values are the means from the three sectors and the errorbars are the standard deviations of the quantities.

\begin{table} \centering
\caption{\em Summary of the parameters for the WD code solution of the \tess\ light curve
of binary~B. The quoted values and errorbars are the mean and standard deviation of the
values from the three sectors for each parameter. Descriptions of the control and fitting
parameters of the WD code can be found in Ref.~\cite{WilsonVanhamme04}. \label{tab:wd2}}
\begin{tabular}{lcc}
{\em Parameter}                                      & {\em Star Ba    }     & {\em Star Bb}         \\[3pt]           
{\it Control parameters:} \\                                                                                           %
{\sc wd2004} operation mode                          & \multicolumn{2}{c}{2}                         \\                
Treatment of reflection                              & \multicolumn{2}{c}{1}                         \\                
Number of reflections                                & \multicolumn{2}{c}{1}                         \\                
Limb darkening law                                   & \multicolumn{2}{c}{1 (linear)}                \\                
Numerical grid size (normal)                         & \multicolumn{2}{c}{50}                        \\                
Numerical grid size (coarse)                         & \multicolumn{2}{c}{40}                        \\[3pt]           
{\it Fixed parameters:} \\                                                                                             %
Orbital eccentricity                                 & \multicolumn{2}{c}{0}                         \\                
Rotation rates                                       & 1.0                   & 1.0                   \\                
Gravity darkening                                    & 1.0                   & 1.0                   \\                
\Teff\ (K)                                           & 20\,000               &                       \\                
Bolometric limb darkening coefficient                & 0.700                 & 0.710                 \\                
Linear limb darkening coefficient                    & 0.233                 & 0.245                 \\                
Bolometric albedos                                   & 1.0                   & 1.0                   \\[3pt]           
{\it Fitted parameters:} \\                                                                                            %
Potential                                            & $3.217 \pm 0.020$     & $3.25  \pm 0.11$      \\                
Mass ratio                                           & \multicolumn{2}{c}{$0.542\pm 0.025$}          \\                
Orbital inclination (\degr)                          & \multicolumn{2}{c}{$88.1 \pm 1.2$}            \\                
Third light                                          & \multicolumn{2}{c}{$0.8785 \pm 0.0039$}       \\                
Light contribution                                   & $1.083 \pm 0.053$     &                       \\[3pt]           
\Teff\ (K)                                           &                       & $16100 \pm 1800$      \\                
{\it Derived parameters:} \\                                                                                           %
Fractional radii                                     & $0.3860 \pm 0.0033$   & $0.2743 \pm 0.0069$   \\[10pt]          
\end{tabular}
\end{table}


\section*{Measurement of the orbital periods}

\begin{figure}[t] \centering \includegraphics[width=\textwidth]{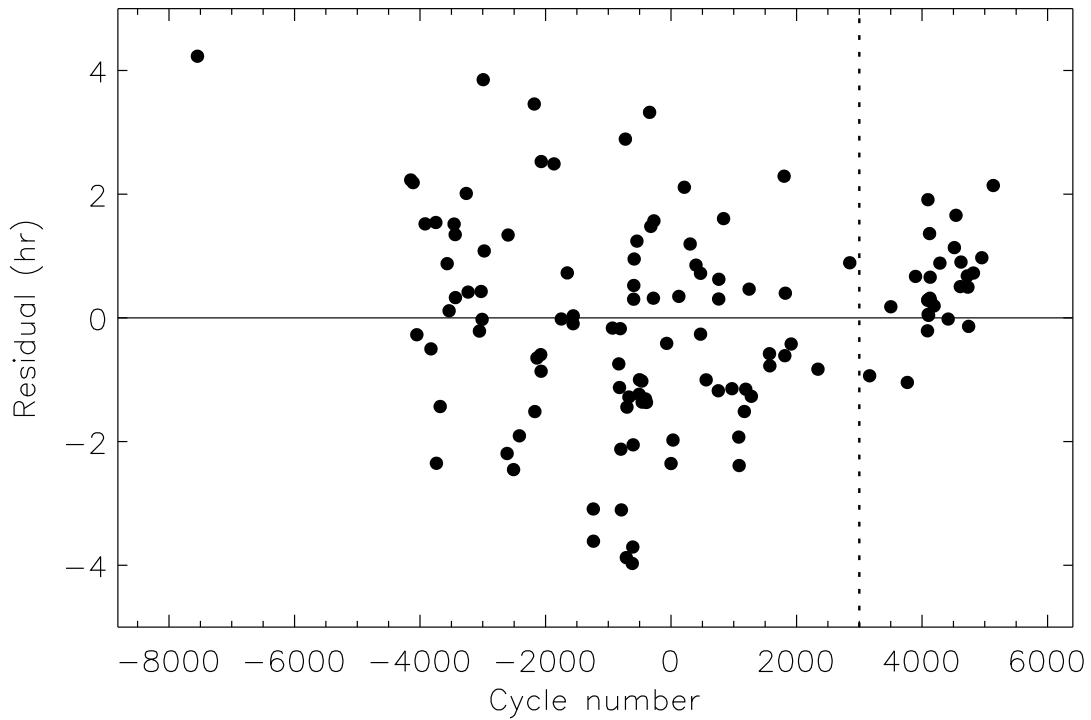} \\
\caption{\label{fig:minall} Residuals of the fit of a linear ephemeris to the
historical times of minimum light recorded for V498~Cyg. The dotted line
divides the older timings obtained using visual, photographic and photoelectric
methods from the more recent CCD observations. Note the size of y-axis scale.} \end{figure}

Eclipse timing measurements of V498~Cyg can be traced back over a century, beginning with times of photographic plates when the system was noticed to be unusually faint, progressing through the eras of detailed study via photographic and photoelectric methods, and continuing into current times with CCD minima primarily obtained by amateur astronomers. An extensive set of eclipse timing measurements was kindly made available by Dr.\ Jerzy Kreiner (see Ref.\ \cite{Kreiner++01book}) to which we attempted to fit a linear ephemeris.

The results are shown in Fig.~\ref{fig:minall} and are discouraging: the scatter around the best fit is huge for the older timings (as expected), but also for the recent CCD timings. Scatters of order minutes are expected from CCD observations of systems with the brightness and eclipse depth of V498~Cyg, but instead a scatter of several hours is seen. We suspect that the presence of the eclipses from binary~B has deleteriously affected many or most of the eclipse timings for binary~A; other explanations are the existence of dynamical effects between the two dEBs if they are gravitationally bound, and that the eclipses of binary~A last approximately 15~hr so are too long to fit into an observing night for a single telescope. For the record, the ephemeris obtained from the fit in Fig.~\ref{fig:minall} is
\begin{equation}
\mbox{Min~I} = \mbox{BJD/TDB } 2440007.6116 + 3.4848627 E
\end{equation}
with a root-mean-square scatter of 5730~s (i.e.\ 1.6~hr). This fit was performed without weighting the data because most of the eclipse timings have no quoted uncertainty and for those that do they are negligible compared to the scatter. Uncertainties in the ephemeris are not given for the same reason. The historical timings were all converted to the BJD/TDB timescale using the routines of Eastman \etal\ \cite{Eastman++10pasp} before fitting the ephemeris.

\begin{figure}[t] \centering \includegraphics[width=\textwidth]{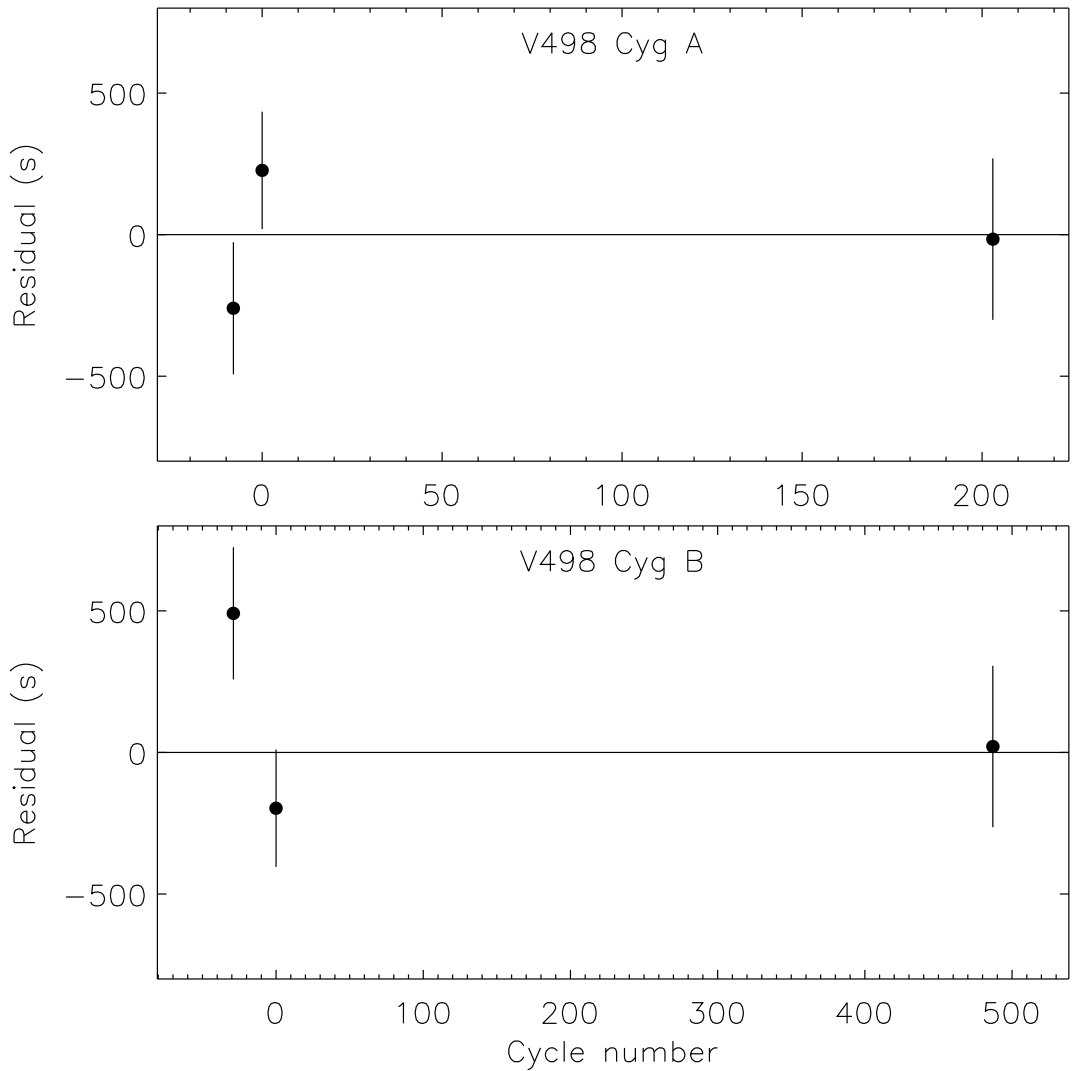} \\
\caption{\label{fig:minAB} Residuals of the fits of linear ephemerides to the times of
minimum light measured for binary~A (top) and binary~B (bottom) in this work.} \end{figure}

Considering the difficulties found above, and the lack of published eclipse timings for binary~B, we instead determined precise orbital periods for the two dEBs using only our fits to the three sectors of \tess\ observations. A linear ephemeris was adopted for each system and the uncertainties in the eclipse timings were scaled to yield $\chi^2_\nu = 1$ for each dEB. The timings are given in Table~\ref{tab:timings} and the fits are shown in Fig.~\ref{fig:minAB}. The ensuing ephemerides are
\begin{equation}
\mbox{Min~I} = \mbox{BJD/TDB } 2458723.1091 (16) + 3.484806 (16) E
\end{equation}
for binary~A and
\begin{equation}
\mbox{Min~I} = \mbox{BJD/TDB } 2458727.4821 (20) + 1.438457 (8) E
\end{equation}
for binary~B, where a bracketed quantity indicates the uncertainties in the final digit of the immediately preceding number. The residuals of the fits to the ephemerides of the two components are anti-correlated, suggesting the presence of mutual dynamical interactions between two gravitationally-bound binary systems. If the interpretation is correct, V498~Cyg is a particularly interesting system for future study.


\section*{Discussion and conclusions}

V498~Cyg has been known to be eclipsing for over 80~yr. The \tess\ light curve reveals a second set of eclipses which are much shallower and have a shorter orbital period (3.48~d for binary~A and 1.44~d for binary~B). These data have been fitted using a model developed specifically for this system, {\sc multebop}, which generates light curves for the two dEBs individually using the {\sc jktebop} code and combines them alongside normalisation polynomials and third light. This allowed all relevant parameters to be fitted together to give the overall best model.

We were, however, unable to get a good model for the system using this approach. The residuals of the fit (Fig.~\ref{fig:resid}) are large, depend primarily on the orbital phase of binary~A, and change during and between the time intervals covered by the data from the three \tess\ sectors. A visualisation of the best fit for binary~A (Fig.~\ref{fig:phase1}) shows a poor agreement with the shapes of the eclipses, most obviously an inability to correctly reproduce the asymmetric brightness variation at the points of first and fourth contact in the primary eclipse. There is a clear perturbation to the light curve which may be due to pulsations, surface inhomogeneities, and/or mass transfer processes.

The large fractional radii of all four stars prompted a separate examination of the light curves of the two dEBs using the WD code. A good fit to binary~B was obtained and some of its photometric properties were constrained to reasonable precision. Binary A was more challenging and an acceptable model of the system was not found.

\begin{table} \centering
\caption{\em \label{tab:timings} Best-fitting parameters and uncertainties from the
{\sc multebop} fit to the three sectors of data. The parameter values are the means
from fits to the three sectors individually. The uncertainties are the standard
deviations (not standard errors) of the fits to six half-sectors individually.}
\begin{tabular}{cr@{\,$\pm$\,}lr@{\,$\pm$\,}l}
{\em TESS sector} & \multicolumn{2}{c}{\em Binary A} & \multicolumn{2}{c}{\em Binary B} \\[3pt]
14                &   2458695.22760 & 0.0024         &     2458685.7726 & 0.0039        \\
15                &   2458723.11168 & 0.0022         &     2458727.4799 & 0.0024        \\
41                &   2459430.52441 & 0.0030         &     2459428.0112 & 0.0033        \\
\end{tabular}
\end{table}

We can nevertheless construct a straw-man model of the system using the constraints we have assembled, beginning with binary~A. Lacy \cite{Lacy90ibvs} found its spectral type to be B1.5~V, and the depths and widths of the spectral lines of the two components to be similar. Adopting an effective temperature of 25\,000~K for star~Aa \cite{PecautMamajek13apjs} and a zero-age main sequence isochrone for solar metallicity from the PARSEC models \cite{Bressan+12mn}, we infer a mass of 10\Msun\ and a radius of 5\Rsun. Star~Ab appears to be more evolved than star~Aa, based on our measurement that its temperature is lower (21\,000~K from the surface brightness ratio in Table~\ref{tab:lc}) but its radius is larger (6\Rsun), so our assumption of zero age is questionable. Given the difficulties found in fitting the light curve, we regard these properties (especially for star~Ab) as unreliable. The system is expected to be young because it has a significant orbital eccentricity despite the short tidal circularisation timescale for stars of this mass and fractional radius \cite{Zahn77aa}, although an alternative explanation is that a small orbital eccentricity is maintained by dynamical effects if the two dEBs are gravitationally bound.

Turning to binary~B, a simple scaling of its luminosity according to the light ratio between the dEBs suggests masses and radii of roughly 6.5\Msun\ and 3.3\Rsun, respectively, for star~Ba. The mass and radius ratios from the fit with the WD code then give the equivalent properties for star~Bb to be 3.5\Msun\ and 2.3\Rsun, which is an acceptable match to the predictions from the PARSEC models. The predicted tempertures for these masses on the ZAMS are approximately 20\,000~K and 14\,000~K, a ratio which is only mildly discrepant with the WD code fit. The estimated properties of the four stars are collected in Table~\ref{tab:absdim} for reference.

Using Kepler's third law we infer that the semimajor axes of the relative orbits of the two dEBs are 0.13~au and 0.05~au. The fractional radii implied by this and the radii suggested above are all significantly lower than those actually measured. We conclude that the system has either experienced significant evolution and is beyond the zero-age main sequence, mass transfer has made the stars unrepresentative of single-star evolution, or the masses and thus orbital separations are lower than we have estimated. The straw-man model constructed above is thus unreliable. The small orbital eccentricity detected in binary~A could be the result of dynamical interactions between the two dEBs. These suggestions could be refined by a more sophisticated attempt to match the measured properties of the system to theoretical models, but such an activity would have to carefully navigate the possible binary interactions suggested by the asymmetric primary eclipse in binary~A. A more direct approach is possible and preferable: the system is known to show spectral lines of at least two components so determination of the spectroscopic orbits and thus masses of the two brighter stars is within reach. High-quality observations and careful analysis might also allow the spectral signatures of stars Ba and Bb to be identified and measured. Finally, one more sector of observations with \tess\ are scheduled in August 2022. It will be interesting to see whether this light curve exhibits the same eclipse shapes as those used in the current study, and whether the new eclipse timings reveal clearer evidence of dynamical interactions between the two components of the doubly-eclipsing quadruple star system V498~Cygni.


\begin{table} \centering
\caption{\em \label{tab:absdim} Inferred properties of the four stars in the V498~Cyg system.}
\begin{tabular}{lcccc}
{\em Property} & \mc{\em Binary A} & \mc{\em Binary B} \\
~ & {\em Star Aa} & {\em Star Ab} & {\em Star Ba} & {\em Star Bb} \\
Mass ($\!\!$\Msun) & 10 & 11 & 6.5 & 3.5 \\
Radius ($\!\!$\Rsun) & 5 & 6 & 3.3 & 2.3 \\
Temperature (K) & 25000 & 21000 & 20000 & 14000 \\
\logg\ (log c.g.s) & 4.0 & 3.9 & 4.2 & 4.3 \\
\end{tabular}
\end{table}


\section*{Acknowledgements}

We thank Dr.\ Jerzy Kreiner and Dr.\ Tam\'as Borkovits for helpful discussions during the analysis presented above, and the anonymous referee for a very helpful and exceptionally fast report.
This paper includes data collected by the \tess\ mission. Funding for the \tess\ mission is provided by the NASA's Science Mission Directorate.
The following resources were used in the course of this work: the NASA Astrophysics Data System; the SIMBAD database operated at CDS, Strasbourg, France; and the ar$\chi$iv scientific paper preprint service operated by Cornell University.



\end{document}